\documentclass[12pt,onecolumn,draftclsnofoot]{IEEEtran}

\usepackage{amsmath,amssymb,epsfig,psfrag,cite,subfigure}
\usepackage{dblfloatfix}    % To enable figures at the bottom of page
\usepackage{graphicx}
\usepackage[font=footnotesize]{caption}
\usepackage{algorithm}
\usepackage{epsfig,psfrag}
\usepackage{subfigure}
\usepackage{color}
\usepackage{url}
\usepackage[margin=1 in]{geometry}

\usepackage{algorithm}
\usepackage{algpseudocode}
\usepackage{placeins}

\newcommand{\NL}{{N_{\rm{L}}}}

\newcommand{\mi}{{\rm{m}}_i}

\newcommand{\lT}{{\boldsymbol{l}}_{\rm{T}}}
\newcommand{\lR}{{\boldsymbol{l}}_{\rm{R}}}
\newcommand{\lRB}{\bar{{\boldsymbol{l}}}_{\rm{R}}}

\newcommand{\PTi}{P_{{\rm{T}},i}}
\newcommand{\PTj}{P_{{\rm{T}},j}}

\newcommand{\PRi}{P_{{\rm{R}},i}}
\newcommand{\PRj}{P_{{\rm{R}},j}}

\newcommand{\PRone}{P_{{\rm{R}},1}}
\newcommand{\PRlast}{P_{{\rm{R}},\NL}}
\newcommand{\PR}{{\boldsymbol{P}}_{{\rm{R}}}}
\newcommand{\nT}{{\boldsymbol{n}}_{{\rm{T}}}}
\newcommand{\nR}{{\boldsymbol{n}}_{{\rm{R}}}}

\newcommand{\mtL}{{\mathcal{L}}}

\newcommand{\Ar}{{A_{\mathrm{R}}}}
\newcommand{\norm}[1]{\big\lVert#1\big\rVert}

\newcommand{\LhatR}{{\boldsymbol{\hat{l}}}_{{\rm{R}}}}
\newcommand{\I}{\boldsymbol{I}}
\newcommand{\Thetab}{\boldsymbol{\Theta}}
\newcommand{\LhatRM}{{\boldsymbol{\hat{l}}}_{{\rm{R,MML}}}}

\renewcommand{\baselinestretch}{1.7}

\begin{document}

\include{macros}

\title{Visible Light Positioning under Luminous Flux Degradation of LEDs}

%\author{\IEEEauthorblockN{Furkan Kokdogan and Sinan Gezici}
%\IEEEauthorblockA{\textit{Department of Electrical and Electronics Engineering, Bilkent University,} Ankara, Turkey
%\\
%Emails: \{kokdogan,\,gezici\}@ee.bilkent.edu.tr}
%}

\author{Issifu Iddrisu and\thanks{I. Iddrisu and S. Gezici are with the Department of Electrical and Electronics Engineering, Bilkent University, Ankara 06800, Turkey (e-mails: issifu.iddrisu@bilkent.edu.tr,\,gezici@ee.bilkent.edu.tr). S. Gezici is the \textit{corresponding author}.} Sinan Gezici, \emph{Senior Member, IEEE}}

% make the title area
\maketitle

\vspace{-2cm}

\begin{abstract}
The position estimation problem based on received power measurements is investigated for visible light systems in the presence of luminous flux degradation of light emitting diodes (LEDs). When the receiver is unaware of this degradation and performs position estimation accordingly, there exists a mismatch between the true model and the assumed model. For this scenario, the misspecified Cramér-Rao bound (MCRB) and the mismatched maximum likelihood (MML) estimator are derived to quantify the performance loss due to this model mismatch. Also, the Cramér-Rao lower bound (CRB) and the maximum likelihood (ML) estimator are derived when the receiver knows the degradation formula for the LEDs but does not know the decay rate parameter in that formula. In addition, in the presence of full knowledge about the degradation formula and the decay rate parameters, the CRB and the ML estimator are obtained to specify the best achievable performance. By evaluating the theoretical limits and the estimators in these three scenarios, we reveal the effects of the information about the LED degradation model and the decay rate parameters on position estimation performance. It is shown that the model mismatch can result in significant degradation in localization performance at high signal-to-noise ratios, which can be compensated by conducting joint position and decay rate parameter estimation.

\end{abstract}

\section{Introduction}\label{sec:Intro}

Accurate positioning is crucial in a broad range of indoor applications such as robot navigation, location-aware services, object tracking, and autonomous vehicles \cite{Monica_UWB,EnLighting,Dimian2017,TDOA_VLP_Automative,Platoon_VLP,roadmap}.
Despite extensive research spanning decades on radio frequency (RF) based indoor positioning systems, there remains a lack of a system that is affordable, accurate, and universally accessible. This inherent challenge in RF-based systems stems from multipath propagation complicating the determination of distance or direction related parameters reliably \cite{roadmap,Kbayer_2018,Amanor2018}. Visible light positioning (VLP),  leveraging the widespread adoption of light-emitting diodes (LEDs) as indoor lighting sources, has showcased its potential for accurate indoor localization (\cite{PIEEE,SurveyVLPprocIEEE,Yazar_vlp2020} and references therein). Via VLP systems, accurate localization can be achieved through the effective utilization of existing LED lighting infrastructures, resulting in minimal deployment requirements and energy costs.

In the literature, various position estimation algorithms have been devised, and thorough examinations of theoretical accuracy limits have been conducted for VLP systems. The developed algorithms are based on either a single parameter type such as received signal strength (RSS), time of arrival (TOA), angle of arrival (AOA), and time-difference-of-arrival (TDOA), or a combination of multiple parameters \cite{Steendam2016, sahin2015, Direct_TCOM, Steendam2017,Indoor_OFDM_2016,onTheFundamental,SteendamAOA2017,Yang2014,Prince2015,zhang2014asynchronous,EPSILON,Hua_FusionVLP,Soner_2021,spoVlp,VLP_DP_ICASSP,Zhou_SPO2018}. In practical VLP systems, the use of RSS-based positioning is widespread due to its cost-effective hardware implementation with high positioning accuracy \cite{PIEEE}. Hence, it is commonly preferred over TOA-based techniques, which require synchronization and accurate time delay estimation \cite{zhang2014asynchronous,EPSILON}. In \cite{sahin2015}, both RSS and AOA information is utilized in a visible light communication (VLC) system with multiple LED transmitters and a single receiver for three-dimensional (3D) localization and the derived Cramér-Rao lower bound (CRB) based on the RSS information encompasses 3D positioning across various deployment scenarios. RSS and AOA are also utilized in a 3D indoor positioning approach using a single LED array and multiple VLC receivers \cite{Yang2014}. In \cite{Steendam2016} and \cite{Steendam2017}, two-dimensional VLP utilizing RSS measurements is studied assuming a fixed receiver height. This is achieved through an aperture-based receiver, and the CRB is derived to evaluate the performance of positioning algorithms based on the considered settings.
Also, in \cite{onTheFundamental}, the authors investigate the impact of non-line-of-sight (NLOS) signal components, signal-to-noise ratio (SNR), transmission distance, and the number of LED sources on performance of RSS-based VLP systems, and obtain the closed-form CRB expressions for both location and orientation estimation. In \cite{Direct_TCOM}, a TOA-based CRB is derived to assess the performance of the designed maximum likelihood (ML) estimators employing direct and two-step positioning techniques in synchronous and asynchronous VLP networks. In \cite{spoVlp}, simultaneous position and orientation estimation using multiple LED transmitters and multiple VLC receivers is investigated for RSS-based VLP systems.

In addition to VLP algorithms that are based on visible light propagation models (e.g., \cite{Steendam2016, sahin2015, Direct_TCOM,Steendam2017,Indoor_OFDM_2016,onTheFundamental,SteendamAOA2017,Yang2014,Prince2015,zhang2014asynchronous,EPSILON,spoVlp,VLP_DP_ICASSP}), there also exist data driven (learning) approaches to VLP such as \cite{Chen_ELM, Tran_WOKNN, Arfaoui_DL_2021, van2017weighted,Xu_KNN,Alam_RegFingerprint,huang2017artificial,Yuan_tiltedRx2018,Saadi_kmeans, saadi2019visible, Gradim2018, Guo2017multipleclassifiers, Guo2019two_layerfusion, Liu_singleLEDML,Bakar_multiplePD_ML,Li_ML_VLP,Cao_DL2023}. In \cite{Chen_ELM}, the authors propose an indoor real-time 3D VLP system, employing fingerprinting and the extreme learning machine approach to attain precise positioning, enhanced resistance to interference, and reliable real-time performance. The VLP system initially splits the large indoor positioning environment into regular VLP kernels, thereby decreasing the size of the fingerprint database and leading to a significant exponential reduction in training time. In \cite{Arfaoui_DL_2021}, the authors introduce deep artificial neural network (ANN) models to jointly estimate the 3D position and orientation of a randomly located VLC-receiver, considering random orientation and unknown emitting power. Leveraging RSS-based fingerprinting, the approach involves creating a dataset with instantaneous received SNR, coupled with corresponding 3D position and orientation angles. Also, the study in \cite{Tran_WOKNN} introduces a technique called maximum received signal strength recognition (MRR) and a hybrid algorithm named weighted optimum $k$-nearest neighbors (WOKNN), which a combination of optimum $k$-NN (OKNN) and weighted $k$-NN (WKNN). It is employed to estimate the 2D location of a VLC-receiver using RSS measurements. Although data driven VLP algorithms can provide position estimation in the absence of statistical models, they require collection of training data from the environment, which may be costly and inappropriate for certain applications. In this manuscript, we focus on a model based VLP approach. 

Most of the studies in the VLP literature assume constant power outputs for LED transmitters. However, in practice, luminous flux of LEDs (hence, LED transmit powers) degrade gradually over time \cite{TM21-11}. Although there exist a few studies such as \cite{plets2019impact,SPO_unknownPow2019, PowerUncertainty2022} that take into account the impact of LED transmitter power uncertainty on the accuracy of a VLP system, there is no work in the literature that investigates the effects of luminous flux degradation of LEDs on the accuracy of VLP systems. In this manuscript, we investigate VLP by considering the luminous flux degradation of LEDs. In particular, we quantify the deterioration in localization performance resulting from the mismatch between an ideal model with no decay in power outputs over time and a practical model with certain decay rates by conducting a misspecified Cramér-Rao bound (MCRB) analysis \cite{fortunati2017}. We also derive CRBs and position estimators in the presence of model knowledge related to luminous flux degradation of LEDs. Although an MCRB analysis is conducted for localization in \cite{Cuneyd}, where the authors investigate reconfigurable intelligent surface (RIS) assisted near-field localization for a single-antenna user equipment and single-antenna base station in the presence of line-of-sight blockage with a practical RIS amplitude model, there exist no studies in the literature that derive the MCRB for VLP systems by considering the power decays of LED transmitters. The primary contributions and novel aspects of this study can be outlined as follows:
\begin{itemize}
\item Position estimation problems in visible light systems under luminous flux degradation of LEDs are formulated for the first time in the literature.
\item A mismatched maximum likelihood (MML) estimator is derived for the scenario where the VLC receiver is unaware of the power degradation of the LED transmitters and the corresponding  MCRB and lower bound (LB) expressions are derived and used as benchmarks to assess localization performance in the presence of model mismatch.
\item ML estimators and corresponding CRBs are derived when the VLC receiver knows the power decay formulas for the LEDs by considering two different scenarios with known and unknown decay rate parameters.
\end{itemize}
Also, simulations are conducted to investigate the effects of luminous flux degradation of LEDs by evaluating the proposed estimators and lower bounds in various situations. 

The remainder of the manuscript is organized as follows: In Section~\ref{sec:SysModel}, the system model and the problem formulation are described by specifying the three scenarios considered in the manuscript. In Section~\ref{sec:Sce1}, we consider the scenario in which the VLC receiver is unaware of the degradation in LEDs, and derive the MCRB, the LB, and the MML estimator. Then, the model knowledge is assumed in Section~\ref{sec:Sce2}, and the CRB and the ML estimator are derived for the problem of joint position and decay rate parameter estimation. In Section~\ref{sec:Sce3}, we focus on the scenario with full knowledge of both the degradation model and the decay rate parameter and derive the CRB and the ML estimator to provide a performance benchmark. Various simulation results are presented in Section~\ref{sec:Simulations}, and the concluding remarks are made in Section~\ref{sec:Conc}.

%=========================================================================================

\section{System Model and Problem Description}\label{sec:SysModel}

Consider a VLP system with $\NL$ LED transmitters at known locations denoted by $\lT^i$ for $i\in\{1,\ldots,\NL\}$ and a VLC receiver at an unknown location $\lR$. The VLC receiver is to estimate its location based on signals coming from the LED transmitters. For this purpose, it collects power measurements from the LED transmitters, which are given by \cite{Erdal_CL_2015}
\begin{gather}\label{eq:powMeas}
\PRi=R_p\,\PTi\,h_i(\lR)+\eta_i
\end{gather}
for $i=1,\ldots,\NL$. In \eqref{eq:powMeas}, $\PRi$ is the received power at the VLC receiver due to the signal from the $i$th LED transmitter, $R_p$ represents the responsivity of the photo-detector (PD) at the VLC receiver, $\PTi$ is the transmit power of the $i$th LED transmitter, $h_i(\lR)$ denotes the channel coefficient between the $i$th LED transmitter and the VLC receiver, and $\eta_i$ is zero-mean Gaussian noise with a variance of $\sigma_i^2$, which is independent of $\eta_j$ for all $j\ne i$ \cite{Erdal_CL_2015}. 
As noted from \eqref{eq:powMeas}, it is assumed that the signals coming from different LED transmitters are processed separately and their power levels are measured individually at the VLC receiver. This is based on the use of a certain type of multiple access protocol, such as frequency-division or time-division multiple access \cite{VLP_power_allocation,VLC_Survey,multiaccessVLP}. In addition, the set of possible locations for the VLC receiver is denoted by $\mtL$ (e.g., all possible locations in a given room), that is, $\lR\in\mtL$, and it is assumed that no prior statistical information is available about $\lR$ \cite{Furkan2023}.

As in \cite{PIEEE,CRB_TOA_VLC,sahin2015}, a line-of-sight scenario is considered between each LED transmitter and the VLC receiver, and the channel coefficients in \eqref{eq:powMeas} are calculated as
\begin{gather}\label{eq:alpha}
h_i(\lR)= \frac{(\mi+1) \Ar\left[(\lR - \lT^i)^{T} \nT^i \right]^{\mi} (\lT^i-\lR)^{T} \nR}{2\pi\norm{\lR - \lT^i}^{\mi+3}}
\end{gather}
where $\mi$ is the Lambertian order for the $i$th LED transmitter, $\Ar$ is the area of the PD at the VLC receiver, and $\nR$ and $\nT^i$ represent the orientation vectors of the VLC receiver and the $i$th LED transmitter, respectively \cite{WirelessInfComm_97}. The VLC receiver is assumed to have the knowledge of the parameters $\Ar$, $R_p$, $\nR$, $\mi$, $\lT^i$, and $\nT^i$, and $\sigma_i^2$ \cite{Direct_TCOM,sahin2015}. For instance, the orientation of the VLC receiver ($\nR$) can be determined  by  a  gyroscope  and  the LED parameters  ($\mi$, $\lT^i$, and $\nT^i$)  can be reported to the VLC receiver via visible light communications \cite{Direct_TCOM}.

In practice, luminous flux of an LED gradually depreciates over time, which is commonly modeled by an exponential decay formula as in \cite[Sec.~5.2.4]{TM21-11}. As the luminous flux is directly proportional to the transmitted optical power from an LED through luminous efficacy \cite{LED_Book}, the transmitted power of the $i$th LED in \eqref{eq:powMeas} can be modeled as
\begin{gather}\label{eq:PowDecay}
\PTi=\PTi^0 \, e^{-\alpha_i t}
\end{gather}
for $i\in\{1,\ldots,\NL\}$, where $\PTi^0$ represents the initial power level of the $i$th LED, $t$ is the operating time in hours, and $\alpha_i>0$ is the decay rate for the $i$th LED \cite[Sec.~5.2.4]{TM21-11}. It is assumed that the initial power levels of the LEDs and the operating times are known. (The operating times are taken to be the same without loss of generality.)

Our aim is to analyze the effects of luminous flux degradation of LEDs on visible light positioning based on power measurements. To this aim, we consider three different scenarios:

\textbf{Scenario~1:} The VLC receiver is unaware of the degradation in LEDs, and assumes that the transmit powers are given by $\PTi^0$ for $i=1,\ldots,\NL$. (This is equivalent to assuming that there is no power decay in the LEDs; i.e., $\alpha_i=0$ in \eqref{eq:PowDecay} for all $i\in\{1,\ldots,\NL\}$.)

\textbf{Scenario~2:} The VLC receiver knows the depreciation formula in \eqref{eq:PowDecay}; however, it does not know the decay rates, $\alpha_i$'s.

\textbf{Scenario~3:} The VLC receiver knows the depreciation formula in \eqref{eq:PowDecay} and the decay rates, $\alpha_i$'s.

In the following sections, we derive theoretical limits on position estimation accuracy and obtain estimators that asymptotically achieve those theoretical limits in these three scenarios. By comparing the performance of the estimators and the theoretical limits, the effects of degradation in LEDs can be quantified for various scenarios.

%For the $\NL$ measurements in \eqref{eq:powMeas}, the likelihood function for the position of the VLC receiver can be expressed as
%\begin{align}\label{eq:likelihood}
%p(\PR\,|\,\lR)=
%%\prod_{i=1}^{\NL}\frac{1}{\sqrt{2\pi}\sigma_i}e^{-\frac{(\PRi-\PTi R_p h_i(\lR))^2}{2\sigma_i^2}}
%\left(2\pi\sigma_i^2\right)^{-0.5\NL}e^{-\sum_{i=1}^{\NL}\frac{(\PRi-\PTi R_p h_i(\lR))^2}{2\sigma_i^2}}
%\end{align}

\section{Scenario~1: Model Mismatch}\label{sec:Sce1}

In Scenario~1, the VLC receiver assumes that $\PTi=\PTi^0$ for $i=1,\ldots,\NL$, that is, it is unaware of the degradation in the LEDs. Based on this assumption and the $\NL$ measurements modeled by \eqref{eq:powMeas}, the likelihood function for the position of the VLC receiver can be stated as
\begin{align}\label{eq:likeSce1}
\tilde{p}(\PR\,|\,\lR)=
%\prod_{i=1}^{\NL}\frac{1}{\sqrt{2\pi}\sigma_i}e^{-\frac{(\PRi-\PTi R_p h_i(\lR))^2}{2\sigma_i^2}}
\left(\prod_{i=1}^\NL \frac{1}{\sqrt{2\pi\sigma_i^2}} \right)e^{-\sum\limits_{i=1}^{\NL}\frac{\left(\PRi-\PTi^0 R_p h_i(\lR)\right)^2}{2\sigma_i^2}}
\end{align} 
where $\PR$ represents a vector consisting of the power 
measurements; i.e., $\PR=[ \PRone\cdots\PRlast ]^T$. The expression in \eqref{eq:likeSce1} can be referred to as the \textit{misspecified parametric PDF for $\lR$} since the transmit powers are not specified accurately. 

For comparison purposes, we can specify the true model, which takes into account the luminous flux degradation of the LEDs with time. In this case, the VLC receiver would perfectly know the depreciation formula and the decay rates of the LEDs. Hence, the \textit{true PDF} can be expressed as
\begin{align}\label{eq:likeSce1_2}
p(\PR)=
\left(\prod_{i=1}^\NL \frac{1}{\sqrt{2\pi\sigma_i^2}}\right) e^{-\sum\limits_{i=1}^{\NL}\frac{\left(\PRi-\PTi^0 \, e^{-\alpha_i t} R_p h_i(\lRB)\right)^2}{2\sigma_i^2}}
\end{align}
where $\lRB$ denotes the true position vector of the VLC receiver. 

In order to reveal the degradation caused by the use of the misspecified model in \eqref{eq:likeSce1} instead of the true model in \eqref{eq:likeSce1_2}, we first derive a position estimator and a theoretical limit on estimation accuracy in the presence of model mismatch. 

\subsection{MML Estimator}

The MML estimator corresponds to the parameter that maximizes the misspecified parametric PDF \cite{fortunati2017}. Therefore, in Scenario~1, the MML estimator can be expressed as 
\begin{align}
\label{eq:misEst}
\LhatRM(\PR) = 
\underset{\lR\in\mtL}{\operatorname{arg\,max}}\;\tilde{p}(\PR\,|\,\lR) .
\end{align}
where $\mtL$ denotes the set of possible positions for the VLC receiver. 
%where $\tilde{p}(\PR\,|\,\lR)$ is as in \eqref{eq:likeSce1}. 
From \eqref{eq:likeSce1} and \eqref{eq:misEst}, the MML estimator based on the received signal power measurements $\PR$ in \eqref{eq:powMeas} can be expressed as
\begin{align}\label{eq:MMLestSce1}
  \LhatRM(\PR) = 
\underset{\lR\in\mtL}{\operatorname{arg\, min}}\;\sum_{i=1}^{\NL}\frac{(\PRi-\PTi^0 \, R_p h_i(\lR))^2}{2\sigma_i^2}\,\cdot
\end{align}
It can be shown that under suitable regularity conditions, the MML estimator is asymptotically unbiased and its error covariance matrix is equal to the MCRB, which is derived next \cite{fortunati2017}. 

%In this section, our objective is to measure the decrease in localization performance caused by the model not accounting for the degradation of the LEDs' power. To achieve our objective, we obtain the misspecified CRB (MCRB) of the model \cite{fortunati2017}, \cite{FORTUNATI2018197}. Subsequent to this, we describe the true model, which conforms to scenario III where the VLC receiver knows depreciation formulation and the decay rates. Then, we derive mismatched maximum likelihood (MML) estimator, and the MCRB and its corresponding lower bound (LB).

\subsection{MCRB}

To derive the theoretical limit on position estimation accuracy in Scenario~1, we first introduce the  pseudo-true parameter \cite{fortunati2017}, which is the parameter that minimizes the Kullback-Leibler (KL) divergence between the true PDF ${p}(\PR)$ and the misspecified parametric PDF $\tilde{p}(\PR\,|\,\lR)$. The pseudo-true parameter can be obtained as follows:
\begin{equation}
\label{eq:pseudoTrue}
    \lR^0 = 
\underset{\lR\in\mathbb{R}^3}{\operatorname{arg\,min}}  \,D\,(p(\PR) \| \tilde{p}(\PR\,|\,\lR))
\end{equation}
where $D\,(p(\PR) \| \tilde{p}(\PR\,|\,\lR))$ denotes the KL divergence between the PDFs ${p}(\PR)$ and $\tilde{p}(\PR\,|\,\lR)$.

Let $\LhatR(\PR)$ denote a misspecified-unbiased (MS-unbiased) estimator of the true position, ${\boldsymbol{\bar{l}}}_R$. That is, the expected value of the estimator $\LhatR(\PR)$ under the true model is equal to $\lR^0$ \cite{Cuneyd}. The MCRB is a lower bound on the covariance matrix of any MS-unbiased estimator $\LhatR(\PR)$ of ${\boldsymbol{\bar{l}}}_R$ and is given by \cite{fortunati2017},\cite{FORTUNATI2018197}
\begin{align}
\mathbb{E}_p\big\{(\LhatR(\PR) - \lR^0)(\LhatR(\PR)- \lR^0 )^T\big\} \succeq \text{MCRB}(\lR^0) 
\end{align}
where ${\mathbb{E}}_p\{\cdot\}$  denotes the expectation under the true model $p(\PR)$, and 
\begin{equation}
\label{eq:mcrb}
\text{MCRB}(\lR^0) \triangleq \mathbf{A}_{\lR^0}^{-1}\mathbf{B}_{\lR^0}\mathbf{A}_{\lR^0}^{-1}.
\end{equation}
The $(m,n)$th elements of the matrices $\mathbf{A}_{\lR^0}$ and $\mathbf{B}_{\lR^0}$ in \eqref{eq:mcrb} are calculated as
\begin{align} 
\label{eq:matrixA}
\big[\mathbf{A}_{\lR^0}\big]_{mn}&=
{\mathbb{E}}_{p}\left\{\frac{\partial^{2}\log \tilde{p}(\PR\,|\,\lR)}{\partial \lR(m)\partial \lR(n)}\bigg{\vert}_{\lR = \lR^0}\right\}, 
%\end{align}
%\begin{align*} 
\\
\big[\mathbf{B}_{\lR^0}\big]_{mn}&=
{\mathbb{E}}_{p} \Bigg\{\frac{\partial \log \tilde{p}(\PR\,|\,\lR)}{\partial \lR(m)}  
\frac{\partial \log \tilde{p}(\PR\,|\,\lR)}{\partial \lR(n)}\bigg{\vert}_{\lR=\lR^0}
\Bigg\}\label{eq:matrixB}
\end{align}
for $1 \leq m,n \leq 3$ with $\lR(n)$ denoting the $n$th element of $\lR$. As we are typically not interested in the value of the pseudo-true parameter, the MCRB is employed to establish a lower bound (LB) for any MS-unbiased estimator with respect to the true parameter value \cite{fortunati2017,Cuneyd}. Namely, the error covariance matrix for any MS-unbiased estimator $\LhatR(\PR)$ is lower bounded as follows:
\begin{align}
{\mathbb{E}}_p\,\big\{(\LhatR(\PR) -\lRB)( \LhatR(\PR)-\lRB )^T\big\} \, \succeq  \, \text{LB}(\lR^0)
\end{align}
where 
\begin{equation} 
\label{eq:lowerBound}
\text{LB}(\lR^0) \triangleq \text{MCRB}(\lR^0) + (\lRB - \lR^0)(\lRB- \lR^0 )^T.
\end{equation}
In \eqref{eq:lowerBound}, the last term is a bias term and it is independent of the signal-to-noise ratio (SNR). As a result, as the SNR goes to infinity, the MCRB goes to zero and the bias term becomes tight with the error covariance matrix of the MS-unbiased estimator \cite{Cuneyd}.

To derive the MCRB, hence, the LB in \eqref{eq:lowerBound}, the following derivations are performed:

\subsubsection{Determination of Pseudo-True Parameter}

To derive the MCRB related to the estimation of the VLC receiver position under the model mismatch, we first need to calculate the pseudo-true parameter $\lR^0$ in \eqref{eq:pseudoTrue}. Specifically, we should find the value of $\lR$ that minimizes the KL divergence between the PDFs ${p}(\PR)$ in \eqref{eq:likeSce1} and $\tilde{p}(\PR\,|\,\lR)$ in \eqref{eq:likeSce1_2}. The value of $\lR$ that minimizes this KL divergence can be derived from \eqref{eq:likeSce1}, \eqref{eq:likeSce1_2}, and \eqref{eq:pseudoTrue}, after some manipulation, as 
\begin{equation}
\label{pseudoTrueParameter}
\lR^0 =
\underset{\lR\in\mathbb{R}^3}{\operatorname{arg\,min}}\sum_{i=1}^{\NL}\frac{(\PTi^0 \, e^{-\alpha_i t} R_p h_i({\boldsymbol{\bar{l}}}_R) -\PTi^0 \, R_p h_i(\lR) )^2}{2\sigma_i^2} 
\end{equation}
The details of the derivation are presented in \textbf{Appendix~\ref{app:Pseudo}}. 

\subsubsection{Derivation of MCRB}

Once the pseudo-true parameter is obtained, we compute the matrices $\mathbf{A}_{\lR^0}$ in \eqref{eq:matrixA} and $\mathbf{B}_{\lR^0}$ in \eqref{eq:matrixB} for calculating the MCRB from \eqref{eq:mcrb}. Based on the expressions in \eqref{eq:likeSce1} and \eqref{eq:likeSce1_2}, the second-order derivatives in \eqref{eq:matrixA} can be obtained as follows:
%\begin{align*} 
%[\mathbf{A}_{\lR^0}]_{mn}=
%E_{p}\left\{\frac{\partial^{2}}{\partial \lR(m)\partial %\lR(n)}\log \tilde{p}(\PR\,|\,\lR)\bigg{\vert}_{\lR = \lR^0}%\right\} 
%\end{align*}
\begin{align}\nonumber
\frac{\partial^{2} \log \tilde{p}(\PR\,|\,\lR)}{\partial \lR(m)\partial \lR(n)} &= 
\sum_{i=1}^{\NL}\frac{(\PRi\,\PTi^0 \, R_p )}{\sigma_i^2} \frac{\partial^{2} h_i(\lR)}{\partial \lR(m)\partial \lR(n)}
\\&-\sum_{i=1}^{\NL}\frac{(\PRi\,\PTi^0  )^2}{\sigma_i^2} \frac{\partial  h_i(\lR) }{\partial \lR(m)} \frac{\partial  h_i(\lR) }{\partial \lR(n)}
\nonumber\\
\label{eq:SecDerA}
&-\sum_{i=1}^{\NL}\frac{(\PRi\,\PTi^0  )^2}{\sigma_i^2} h_i(\lR)\frac{\partial^{2} h_i(\lR)}{\partial \lR(m)\partial \lR(n)}\cdot
\end{align}
Then, by evaluating the expression in \eqref{eq:SecDerA} at $\lR=\lR^0$, the elements of matrix $\mathbf{A}_{\lR^0}$ in \eqref{eq:matrixA} can be derived as 
\begin{align} \nonumber
\big[\mathbf{A}_{\lR^0}\big]_{mn}&=
\sum_{i=1}^{\NL}\frac{{\mathbb{E}}_{p}\{\PRi\}\,\PTi^0 \, R_p }{\sigma_i^2} \frac{\partial^{2} h_i(\lR^0)}{\partial \lR(m)\partial \lR(n)}
\\&-
\sum_{i=1}^{\NL}\frac{(\PTi^0 \,R_p)^2}{\sigma_i^2} \frac{\partial  h_i(\lR^0) }{\partial \lR(m)} \frac{\partial  h_i(\lR^0) }{\partial \lR(n)}
\nonumber\\
\label{eq:matrixAexplicit}
&-\sum_{i=1}^{\NL}\frac{(\PTi^0 \, R_p )^2}{\sigma_i^2} h_i(\lR^0)\frac{\partial^{2} h_i(\lR^0)}{\partial \lR(m)\partial \lR(n)}
\end{align}
where ${\mathbb{E}}_{p}\{\PRi\}  = \PTi^0 \, e^{-\alpha_i t} R_p h_i(\lRB)$.

Similarly, via \eqref{eq:likeSce1} and \eqref{eq:likeSce1_2}, the first-order derivatives in \eqref{eq:matrixB} are calculated as
%\begin{align*} 
%[\mathbf{B}_{\lR^0}]_{mn}=
%E_{p} \left\{\frac{\partial \log \tilde{p}(\PR\,|\,\lR)}%{\partial \lR(m)}\bigg{\vert}_{\lR = \lR^0} \cdot 
%\frac{\partial \log \tilde{p}(\PR\,|\,\lR)}{\partial \lR(n)}%\bigg{\vert}_{\lR=\lR^0}
%\right\}.
%\end{align*}
\begin{align}
\frac{\partial \log \tilde p(\PR\,|\,\lR) }{\partial \lR(m)} = \sum_{i=1}^{\NL} & \PTi^0 \, R_p  \frac{(\PRi-\PTi^0 \, R_p h_i(\lR))}{\sigma_i^2} 
\label{eq:firstDerMatB}
\frac{\partial h_i(\lR)}{\partial \lR(m)}\,\cdot
\end{align}
Then, based on \eqref{eq:firstDerMatB}, the elements of matrix $\mathbf{B}_{\lR^0}$ in \eqref{eq:matrixB} can be obtained as follows:
\begin{align}\nonumber
\big[\mathbf{B}_{\lR^0}\big]_{mn}=&
\,{\mathbb{E}}_{p} \Bigg\{\sum_{i=1}^{\NL} \bigg(\PTi^0 \, R_p\frac{(\PRi-\PTi^0 \, R_p h_i(\lR^0))}{\sigma_i^2} \frac{\partial h_i(\lR^0)}{\partial \lR(m)}\bigg)
\nonumber\\& \times\sum_{j=1}^{\NL} \bigg(\PTj^0 \, R_p
\frac{(\PRj-\PTj^0 \, R_p h_j(\lR^0))}{\sigma_j^2} \frac{\partial h_j(\lR^0)}{\partial \lR(n)}\bigg)
\Bigg\}
\\&=\sum_{i=1}^{\NL} \Bigg(\left(\frac{\PTi^0 \, R_p}{\sigma_i^2} \right)^2
{\mathbb{E}}_{p}\{(\PRi-\PTi^0 \, R_p h_i(\lR^0))^2\}
\frac{\partial h_i(\lR^0)}{\partial \lR(m)}\frac{\partial h_i(\lR^0)}{\partial \lR(n)} \Bigg) \nonumber\\&+\sum_{i=1}^{\NL}\sum_{j=1
,j \neq i}^{\NL} \bigg(\frac{\PTi^0 \,\PTj R_p^2}{\sigma_i^2\sigma_j^2}(\PTi^0 \, e^{-\alpha_i t} R_p h_i(\lRB) -\PTi^0 \, R_p h_i(\lR^0) )\nonumber\\&
\label{eq:matrixBexplicit}
\times (\PTj^0 \, e^{-\alpha_j t} R_p h_j(\lRB) -\PTj^0 \, R_p h_j(\lR^0) )\frac{\partial h_i(\lR^0)}{\partial \lR(m)}\frac{\partial h_j(\lR^0)}{\partial \lR(n)} \bigg)
\end{align}
where ${\mathbb{E}}_{p}\{(\PRi-\PTi^0 \, R_p h_i(\lR^0))^2\} = \sigma_i^2 + (\PTi^0 \, e^{-\alpha_i t} R_p h_i(\lRB) -\PTi^0 \, R_p h_i(\lR^0) )^2$.

The first and second order derivatives of the channel coefficient $h_i(\cdot)$ appearing in \eqref{eq:matrixAexplicit} and \eqref{eq:matrixBexplicit} can  be derived explicitly in terms of the system parameters, as presented in \textbf{Appendix \ref{app:ChanCoefDer}}. Overall, based on \eqref{pseudoTrueParameter}, \eqref{eq:matrixAexplicit}, and \eqref{eq:matrixBexplicit}, the MCRB in \eqref{eq:mcrb} can be calculated, which also leads to the derivation of the LB in \eqref{eq:lowerBound}. Via this theoretical limit, we can evaluate the effects of omitting the power decays of LEDs.

%Pseudo_True Parameter Derivation
% \begin{align}
% D(p(\PR\,|\,\lR,\bal) \| \tilde{p}(\PR\,|\,\lR)) = \int \log \left( { \frac{p(\PR\,|\,\lR,\bal)}{\tilde{p}(\PR\,|\,\lR)} } \right)p(\PR\,|\,\lR,\bal)dP_R
% \end{align}

% \begin{align}
% D(p(\PR\,|\,\lR,\bal) \| \tilde{p}(\PR\,|\,\lR)) = 
% \mathbb{E}_p \bigg[\log \left( { \frac{p(\PR\,|\,\lR,\bal)}{\tilde{p}(\PR\,|\,\lR)} }\right) \bigg]
% \end{align}

% \begin{align}
% \log p(\PR\,|\,\lR,\bal)  - \log\tilde{p}(\PR\,|\,\lR,\bal) 
% = 
% \sum_{i=1}^{\NL}\frac{(\PRi-\PTi^0 \, R_p h_i(\lR))^2}{2\sigma_i^2} - \sum_{j=1}^{\NL}\frac{(P_{R,j}-P_{T,j}^0 \, e^{-\alpha_j t} R_p h_j(\lR))^2}{2\sigma_j^2} 
% \end{align}

% \begin{align}
% \mathbb{E}_p\big[\log p - \log\tilde{p}\big] = \sum_{i=1}^{\NL}\frac{(\PTi^0 \, e^{-\alpha_i t} R_p h_i(\lRB) -\PTi^0 \, R_p h_i(\lR) )^2}{2\sigma_i^2} 
% \end{align}

% \begin{equation}
% \lR^0 =
% \underset{\lR\in\mathbb{R}^3}{\operatorname{argmin}}\sum_{i=1}^{\NL}\frac{(\PTi^0 \, e^{-\alpha_i t} R_p h_i({\boldsymbol{\bar{l}}}_R) -\PTi^0 \, R_p h_i(\lR) )^2}{2\sigma_i^2} 
% \end{equation}

% \begin{align}
% \text{MCRB}(\lR^0) = \mathbf{A}_{\lR^0}^{-1}\mathbf{B}_{\lR^0}\mathbf{A}_{\lR^0}^{-1}
% \end{align}

% \begin{align*}
% \text{MSE}_p((\LhatR)_{MML},\,\lRB) \succeq \text{MCRB}(\lR^0) + (\lRB -\lR^0)(\lRB -\lR^0)^T
% \end{align*}

% where $\text{MSE}_p((\LhatR)_{MML},\,\lRB) =\mathbb{E}_p\,\{((\LhatR -\lRB)( \LhatR-\lRB )^T)\} $

%==========================

\section{Scenario~2: Model Knowledge}\label{sec:Sce2}

In Scenario~2, the VLC receiver knows the formula in \eqref{eq:PowDecay} but it does not know the decay rates, i.e., the $\alpha_i$ parameters. Typically, the same types of LEDs are employed in a given network. Hence, the decay rates can be taken to be the same in practical cases; i.e., $\alpha_i=\alpha$ for $i=1,\ldots,\NL$. (If all the decay rates were different and unknown, accurate localization might not be possible as there would exist $\NL+3$ unknown parameters with no specific relations and $\NL$ measurements. Specifically, all the power measurements would be scaled by unknown and different factors (the $e^{-\alpha_it}$ terms in \eqref{eq:PowDecay}), making localization an ill-posed estimation problem.\footnote{In such a case, one could simply ignore the power decays in the LEDs and assume that $\PTi=\PTi^0$ for $i=1,\ldots,\NL$, as in Scenario~1.}) In this scenario, the likelihood function for the unknown position vector $\lR$ and the common decay parameter $\alpha$ can be obtained from \eqref{eq:powMeas} and \eqref{eq:PowDecay} as
\begin{align}
\label{eq:likeSce2}
p(\PR\,|\,\lR,\alpha) =
%\prod_{i=1}^{\NL}\frac{1}{\sqrt{2\pi}\sigma_i}e^{-\frac{(\PRi-\PTi R_p h_i(\lR))^2}{2\sigma_i^2}}
\left(\prod_{i=1}^\NL \frac{1}{\sqrt{2\pi\sigma_i^2}}\right) 
 e^{-\sum_{i=1}^{\NL}\frac{\left(\PRi-\PTi^0 \, e^{-\alpha t} R_p h_i(\lR)\right)^2}{2\sigma_i^2}}
\end{align}

In order to evaluate the effects of the unknown decay rate of transmit powers on localization accuracy, we aim to derive the ML estimator and the theoretical limit on position estimation errors, namely, the CRB, in Scenario~2. 

%In this scenario, we aim to measure the decrease in localization performance caused by the VLC receiver lack of knowledge of the specific values of $\alpha_i$ parameters of the LEDs. In order to achieve that aim, we employ the CRB analysis of the model. In the following, we obtain the ML estimator of the model under scenario-II and its CRB. 

%\subsection{The ML Estimator}

%The purpose of the VLC receiver is to estimate the location $\lR$ based on the power measurements in \eqref{eq:powMeas}. 

%\subsubsection{Same $\alpha_i$ Parameters Case}
%With the assumption that all the LEDs have the same decay rates, 

The ML estimator for the position of the VLC receiver in Scenario~2 can be stated as \cite{Poor}
\begin{gather}\label{eq:estSce2}
\big(\LhatR ,  \hat{\alpha}\big)= 
\underset{\lR\in \mtL,\,\alpha>0}{\operatorname{arg\,max}}\;p(\PR\,|\,\lR,\alpha)
\end{gather}
where $\mtL$ denotes the set of possible positions for the VLC receiver. Based on \eqref{eq:likeSce2}, the ML estimator in \eqref{eq:estSce2} can be derived, after some manipulation, as 
% \begin{align} \label{}
% \LhatR = 
% \underset{\lR\in\mathbb{R}^3}
% {\operatorname{arg\,max}}\;\left(\prod_{i=1}^\NL \frac{1}{ \sqrt{2\pi\sigma_i^2}}\right) e^{-\sum_{i=1}^{\NL}\frac{(\PRi-\PTi^0 \, e^{-\hat{\alpha}(\lR) t} R_p h_i(\lR))^2}{2\sigma_i^2}}
% \end{align}
%After performing some operations using \eqref{eq:likeSce2} and \eqref{eq:estSce2}, the ML estimator for the position of the VLC receiver can be expressed as follows:
\begin{align}\label{eq:sce2PosEst}
\LhatR = 
\underset{\lR\in\mtL}{\operatorname{arg\,min}}\;\sum_{i=1}^{\NL}\frac{(\PRi-\PTi^0 \, e^{-\hat{\alpha}(\lR) t} R_p h_i(\lR))^2}{2\sigma_i^2}
\end{align}
where $\hat\alpha(\lR)$ is given by
\begin{equation}\label{eq:sce2PosEstlR}
\hat\alpha(\lR) =
\begin{cases}
-\frac{1}{t}\, \log g(\lR), &\text{if}~0 < g(\lR)< 1 \\
0\,, & \text{otherwise} 
\end{cases}
\end{equation}
with $g(\lR)$ being defined as
\begin{equation}
g(\lR)\triangleq \frac{\sum_{i=1}^{\NL} \frac{\PRi\PTi^0 h_i(\lR)}{\sigma_i^2}}{R_p\sum_{i=1}^{\NL} \frac{(\PTi^0 h_i(\lR))^2}{\sigma_i^2}}\,\cdot
\end{equation}
It is noted that the four dimensional search in \eqref{eq:estSce2} can be reduced to three dimensions in \eqref{eq:sce2PosEst} based on the expression in \eqref{eq:sce2PosEstlR}. Hence, the ML estimator in Scenario~2 and the MML estimator in Scenario~1 (see \eqref{eq:MMLestSce1}) have similar levels of complexity.

%\subsubsection{Different $\alpha_i$ Parameters Case}
%When the LEDs have different decay rates the ML estimator can be expressed as:
%\begin{align}\label{eq:estSce22}
%(\LhatR ,  \hat{\bal})= 
%\underset{\lR\in \mathbb{R}^3,\bal\in\mathbb{R}^{N_L}}{\operatorname{arg\,max}}\;p(\PR\,|\,\lR,\bal)
%\end{align}
%where $\bal=[\alpha_1\,\cdots\,\alpha_{\NL}]^T$ and $\lR =[\lR(1) \,\, \lR(2) \,\, \lR(3) ]^T$

% \begin{align} \label{}
% \LhatR = 
% \underset{\lR}
% {\operatorname{argmax}}\;\left(\prod_{i=1}^\NL \frac{1}{ \sqrt{2\pi\sigma_i^2}}\right) e^{-\sum_{i=1}^{\NL}\frac{(\PRi-\PTi^0 \, e^{-\hat{\alpha}_i(\lR) t} R_p h_i(\lR))^2}{2\sigma_i^2}}
% \end{align}
%After performing some operations using \eqref{eq:likeSce2} and \eqref{eq:estSce22}, the ML estimator for the position of the VLC receiver can be expressed as follows:
%\begin{align}\label{eq:sce2PosEst2}
%\LhatR = 
%\underset{\lR\in \mathbb{R}^3}{\operatorname{arg\,min}}%\;\sum_{i=1}^{\NL}\frac{(\PRi-\PTi^0 \, e^{-\hat{\alpha}_i(\lR) %t} R_p h_i(\lR))^2}{2\sigma_i^2}
%\end{align}
%where  $\hat\alpha_i(\lR)$ given in \eqref{eq:sce2PosEst2} is given by 
%\begin{align}
%\hat\alpha_i(\lR) = 
%\begin{cases}
%-\frac{1}{t}\log\left( \frac{\PRi}{\PTi^0 \,R_p h_i(\lR)}\right), & \text{if $0 < \frac{\PRi}{\PTi^0 \,R_p h_i(\lR)} < 1 $ }\\
%            0, & \text{otherwise}
%		 \end{cases}
%\end{align}

To derive the CRB, we first define the vector of unknown parameters $\Thetab$, which consists of the position of the VLC receiver position, $\lR$, and the decay rate parameter, $\alpha$, namely,  $\Thetab \triangleq [\lR^T \; \alpha]^T =[\lR(1) \; \lR(2) \; \lR(3) \; \alpha]^T$. Then, by denoting the likelihood function in \eqref{eq:likeSce2} as $p(\PR\,|\,\Thetab)$, the Fisher information matrix (FIM) can be expressed as \cite{Poor}
\begin{align}\label{eq:FIM2}
\I(\Thetab) = \mathbb{E}\left\{ \left(\nabla_{\Thetab}\log p(\PR\,|\,\Thetab)\right)  \left(\nabla_{\Thetab}\log p(\PR\,|\,\Thetab)\right)^T\right\} 
\end{align}
where $\nabla_{\Thetab}$ denotes the gradient operator, which yields a four-dimensional column vector in this scenario. Based on the FIM, the CRB on the position estimation error can be stated as follows \cite{Poor}:
\begin{align}\label{eq:sce2lowBound}
 \mathbb{E}\left\{{\|\LhatR -\lR \|}^2\right\} \geq \text{trace}\left\{\I(\Thetab)^{-1}_{1:3,1:3}\right\} \triangleq \text{CRB}_{\lR}
\end{align}
where $\LhatR$ denotes any unbiased estimator of $\lR$ in Scenario~2, and $\I(\Thetab)^{-1}_{1:3,1:3}$ corresponds to the upper $3\times3$ block of the inverse FIM. After some manipulation, the elements of the FIM in \eqref{eq:FIM2} can be computed from \eqref{eq:likeSce2} as 
%\begin{align}\label{}
%[\I(\Thetab)]_{m,n} =  \mathbb{E}\bigg\{ \left(\frac{\partial \log p(\PR\,|\,\Thetab) }{\partial \Thetab(m)}\right)  \left(\frac{\partial  \log p(\PR\,|\,\Thetab)}{\partial \Thetab(n)}\right)\bigg\} 
%\end{align}
% \begin{align}\label{}
% log\;p(\PR\,|\,\lR,\bal) =\sum_{i=1}^{\NL}log\bigg(\frac{1}{ \sqrt{2\pi\sigma_i^2}}\bigg) -\sum_{i=1}^{\NL}\frac{(\PRi-\PTi^0 \, e^{-\alpha t} R_p h_i(\lR))^2}{2\sigma_i^2}
% \end{align}
% \begin{align}\label{}
% \frac{\partial \log p(\PR\,|\,\lR,\bal) }{\partial \Thetab(m)}= R_P \sum_{i=1}^{\NL} \PTi^0 \, \frac{(\PRi-\PTi^0 \, e^{-\alpha t} R_p h_i(\lR))}{\sigma_i^2} \frac{\partial (e^{-\alpha t} h_i(\lR)) }{\partial \Thetab(m)}
% \end{align}
\begin{align}\label{eq:FIMder1}
[\I(\Thetab)]_{m,n} =  R_P^2 \sum_{i=1}^{\NL} \bigg(\frac{(\PTi^0)^2 \,}{\sigma_i^2}\frac{\partial  (e^{-\alpha t} h_i(\lR)) }{\partial \Thetab(m)} \frac{\partial  (e^{-\alpha t} h_i(\lR)) }{\partial \Thetab(n)} \bigg)
\end{align}
for $ m,n\in\{1,2,3,4\}$, where $\Thetab(n)$ denotes the $n$th element of $\Thetab$. In particular, $[\I(\Thetab)]_{m,n}$ can derived explicitly as follows:
\begin{align}\label{eq:FIMsce2_1}
&[\I(\Thetab)]_{m,n} =  R_P^2 \sum_{i=1}^{\NL} \frac{(\PTi^0 \, e^{-\alpha t })^2}{\sigma_i^2} \frac{\partial h_i(\lR) }{\partial \lR(m)} \frac{\partial h_i(\lR) }{\partial \lR(n)}
~~{\rm{for}}~m,n \in \{1,2,3\},
\\\label{eq:FIMsce2_2}
&[\I(\Thetab)]_{4,n} =  -t R_P^2 \sum_{i=1}^{\NL} \frac{(\PTi^0 \, e^{-\alpha t })^2 h_i(\lR)}{\sigma_i^2}   \frac{\partial h_i(\lR) }{\partial \lR(n)}
~~{\rm{for}}~n  \in \{1,2,3\},
\\\label{eq:FIMsce2_3}
&[\I(\Thetab)]_{m,4} =  -t R_P^2 \sum_{i=1}^{\NL} \frac{(\PTi^0 \, e^{-\alpha t })^2 h_i(\lR)}{\sigma_i^2}   \frac{\partial h_i(\lR) }{\partial \lR(m)}
~~{\rm{for}}~
m  \in \{1,2,3\},
\\\label{eq:FIMsce2_4}
&[\I(\Thetab)]_{4,4} =  ( t {R_P} e^{-\alpha t })^2  \sum_{i=1}^{\NL} \frac{(\PTi^0 \, h_i(\lR))^2}{\sigma_i^2}\,\cdot
\end{align}
In \eqref{eq:FIMsce2_1}--\eqref{eq:FIMsce2_4}, the partial derivatives of the channel coefficient $h_i(\lR)$ can be evaluated based on the expressions in \textbf{Appendix~\ref{app:ChanCoefDer}}. Consequently, the elements of the FIM, hence, the CRB in \eqref{eq:sce2lowBound}, can be calculated, which specifies the estimation accuracy limit when the VLC receiver is aware of the power decay model but does not know the decay rate.

%==========================================

\section{Scenario~3: Full Knowledge}\label{sec:Sce3}

In Scenario~3, the VLC receiver knows the formula in \eqref{eq:PowDecay} and the decay rate parameters, $\alpha_i$'s. In other words, Scenario~3 corresponds to the scenario with perfect knowledge and provides a performance bound for the other scenarios. Based on \eqref{eq:powMeas} and \eqref{eq:PowDecay}, the likelihood function in Scenario~3 is given by
\begin{align}
\label{eq:likeSce3}
p(\PR\,|\,\lR)&=
%\prod_{i=1}^{\NL}\frac{1}{\sqrt{2\pi}\sigma_i}e^{-\frac{(\PRi-\PTi R_p h_i(\lR))^2}{2\sigma_i^2}}
\left(\prod_{i=1}^\NL \frac{1}{ \sqrt{2\pi\sigma_i^2}} \right)
 e^{-\sum_{i=1}^{\NL}\frac{\left(\PRi-\PTi^0 \, e^{-\alpha_i t} R_p h_i(\lR)\right)^2}{2\sigma_i^2}}\cdot
\end{align}
%In this scenario, we aim to measure the performance of the ML estimator when $\alpha_i$ parameters of the LEDs are perfectly known by the VLC receiver. In order to achieve that aim, we employ the CRB analysis to establish the benchmark for the performance of the ML estimator.In the following, we obtain the ML estimator of the model under scenario-III where the VLC receiver has perfect knowledge of the formula and the $\alpha_i$ parameters and its CRB. 
%Here, the position of the VLC receiver is the only unknown vector parameter and the decay rates $\bal_i$ parameters are assumed to be perfectly known. To estimate the VLC receiver location, we maximize the likelihood function $p(\PR\,|\,\lR)$ from \eqref{eq:likeSce3} over $\lR$. The value of $\lR$ that maximizes the likelihood function $p(\PR\,|\,\lR)$ from \eqref{eq:likeSce3} can be obtained as follows \cite{FORTUNATI2018197}
Accordingly, the ML estimator is expressed as
\begin{align}\label{eq:estSce3}
\LhatR = 
\underset{\lR\in\mtL}{\operatorname{arg\,max}}\;p(\PR\,|\,\lR)
\end{align}
which can be specified, based on \eqref{eq:likeSce3}, as follows:
% \begin{align}\label{}
% \log\;p(\PR\,|\,\lR) =\sum_{i=1}^{\NL}\log\bigg(\frac{1}{ \sqrt{2\pi\sigma_i^2}}\bigg) -\sum_{i=1}^{\NL}\frac{(\PRi-\PTi^0 \, e^{-\alpha_i t} R_p h_i(\lR))^2}{2\sigma_i^2}
% \end{align}
%After performing some operations using \eqref{eq:likeSce3} and \eqref{eq:estSce3}, the ML estimator for the position of the VLC receiver can be expressed as
\begin{align}\label{eq:MLestSce3}
\LhatR = 
\underset{\lR\in\mtL}{\operatorname{arg\,min}}\;\sum_{i=1}^{\NL}\frac{(\PRi-\PTi^0 \, e^{-\alpha_i t} R_p h_i(\lR))^2}{2\sigma_i^2}\cdot
\end{align}
Similar to the ML estimator in Scenario~2 (see \eqref{eq:sce2PosEst} and \eqref{eq:sce2PosEstlR}), the ML estimator in \eqref{eq:MLestSce3} also requires a three-dimensional search. However, it does not need to estimate the decay rate during the search (cf.~\eqref{eq:sce2PosEstlR}) as it is assumed to be known in Scenario~3.

% \begin{figure}
% \includegraphics[width=1\linewidth]{vlp_images/alpha_plot_t11.eps}
% \centering
% \caption{Plot of CRB versus $\alpha$}
% % \label{}
% \end{figure}

To evaluate the CRB in Scenario~3, we first calculate the FIM, which is defined as \cite{Poor}
\begin{align}\label{eq:FIM3}
\I(\lR) = \mathbb{E}\left\{ \left(\nabla_{\lR}\log p(\PR\,|\,\lR)\right)  \left(\nabla_{\lR}\log p(\PR\,|\,\lR)\right)^T\right\} 
\end{align}
with $\nabla_{\lR}$ denoting the gradient operator with respect to $\lR$. Based on the FIM, the CRB on the position estimation error is specified as \cite{Poor}
\begin{align}\label{eq:CRLBsce3}
 \mathbb{E}\left\{{\|\LhatR -\lR \|}^2\right\} \geq  \text{trace}\left\{\I(\lR)^{-1}\right\} \triangleq \text{CRB}
\end{align}
where $\LhatR$ denotes an unbiased estimate of $\lR$ in Scenario~3. After some manipulation, the elements of the FIM in \eqref{eq:FIM3} can be derived from \eqref{eq:likeSce3} as
\begin{align}\label{eq:FIMsce3}
[\I(\lR)]_{m,n} =  R_P^2 \sum_{i=1}^{\NL} \frac{\left(\PTi^0 \, e^{-\alpha_i t}\right)^2}{\sigma_i^2} \frac{\partial h_i(\lR) }{\partial \lR(m)}\frac{\partial h_i(\lR) }{\partial \lR(n)}
\end{align}
for $m,n\in\{1,2,3\}$. The FIM can be evaluated via \eqref{eq:FIMsce3} and the partial derivative expressions in \textbf{Appendix~\ref{app:ChanCoefDer}}, leading to the evaluation of the CRB in \eqref{eq:CRLBsce3}.

% \begin{align}\label{}
% \frac{\partial \log p(\PR\,|\,\lR) }{\partial l_R(m)}= \sum_{i=1}^{\NL} \PTi^0 \, e^{-\alpha_i t}R_P \frac{(\PRi-\PTi^0 \, e^{-\alpha_i t} R_p h_i(\lR))}{2\sigma_i^2} \left(\frac{\partial h_i(\lR) }{\partial l_R(m)}\right)
% \end{align}

% \begin{align}\label{}
% [\I(\lR)]_{m,n} =  {R_P}^2 \sum_{i=1}^{\NL}
% \mathbb{E}\{ (\PRi-\PTi^0 \, e^{-\alpha_i t} R_p h_i(\lR))^2\}\left(\frac{\PTi^0 \, e^{-\alpha_i t}}{\sigma_i^2}\right)^2 \left(\frac{\partial h_i(\lR) }{\partial l_R(m)}\right) \left(\frac{\partial h_i(\lR) }{\partial l_R(n)}\right)
% \end{align}

% $\PRi$ is independent of $\PRj$ for all $j\ne i$

% \begin{align}\label{}
% [\I(\lR)]_{m,n} =  {R_P}^2 \sum_{i=1}^{\NL} \frac{\left(\PTi^0 \, e^{-\alpha_i t}\right)^2}{\sigma_i^2} \left(\frac{\partial h_i(\lR) }{\partial l_R(m)}\right) \left(\frac{\partial h_i(\lR) }{\partial l_R(n)}\right)
% \end{align}

\textbf{Remark:} Based on the MCRB and the LB expressions in Scenario~1 and the CRB expressions in Scenario~2 and Scenario~3, the effects of omitting the power decay of LEDs, considering the decay model, and knowing the decay rate parameters can be evaluated in various situations. Since these theoretical limits are achieved by the MML and the ML estimators at high SNRs, they can be used to assess the localization performance and to design system parameters in high SNR conditions.

\section{Simulation Results}\label{sec:Simulations}

In this section, we conduct simulations to investigate the performance of the proposed approaches and to evaluate the theoretical limits. Similar to \cite{Furkan2023}, we consider a room with dimensions of $4 \times 4 \times 3$ meters (width, depth, and height, respectively) and 9 LEDs  (i.e., $\NL = 9$) placed at the following coordinates in meters: \{(-1, 1, 3), (0, 1, 3), (1, 1, 3), (-1, 0, 3), (0, 0, 3), (1, 0, 3), (-1, -1, 3), (0, -1, 3), (1, -1, 3)\}. These LED positions are selected to ensure a symmetrical coverage of the room, with the center of the room floor being represented by the coordinates (0, 0, 0). The orientations of the LEDs are defined as $\nT^i= [0, 0, -1]^T$ for $i=1,\ldots,\NL$, indicating that they all point downwards. Also, the Lambertian orders of the LEDs, $\mi$'s, are set to 1. The responsivity of the PD is set to $R_p = 1$, and the area of the PD is taken as $\Ar = 1 \,{\mathrm{cm}}^{2}$. The orientation of the receiver is defined as $\nR= [0, 0, 1]^T$, indicating that it points upwards. Furthermore, we assume that the noise variances are equal, denoted as $\sigma_i^2 = \sigma^2$ for $i = 1,\ldots,\NL$.

\begin{figure}
\includegraphics[width=0.95\linewidth]{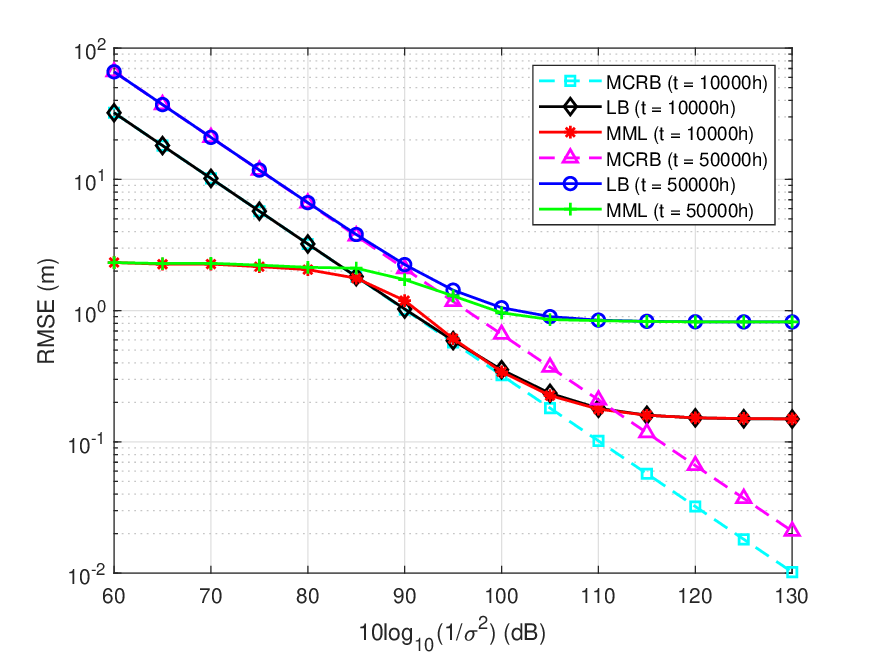}
\centering
\caption{Performance of the MML estimator and the corresponding  MCRB and LB versus $10\log_{10} (1/\sigma^2)$ in Scenario~1 for LED decay rates of $\alpha = 10^{-5}$ and time intervals of $t \in \{10000\,h,\, 50000\,h\}$ when the VLC receiver is positioned at $(0.5,0,5,0.85)\,$m.} 
\label{fig:scen1figure}
\end{figure}

To evaluate the performance of the MML estimator under  Scenario~1 and the ML estimators under  Scenario~2 and  Scenario~3, the VLC receiver is placed at $\lR = (0.5,\, 0.5,\,  0.85)$ meters, the initial power levels of the LEDs are set to be identical as $\PTi^0= 10\,$W, and the decay rates of the LEDs are assumed to be the same and taken as $10^{-5}$ based on the decay rate calculations in \cite[Sec.~7]{TM21-11}. In Fig. \ref{fig:scen1figure}, the RMSE performance of the MML estimator in Scenario~1 and the corresponding MCRB and LB are plotted versus $10\log_{10} (1/\sigma^2)$ for two different time intervals ($10000$ and $50000$ hours). It can be observed from the figure that for both time intervals, the LB plateau at low noise variances and the MCRB goes to zero, as expected. Additionally, it is worth noting that the MCRB and LB values at $t = 10000\,$h are lower than those at $t = 50000\,$h. This is due to the gradual reduction in power output of the LED transmitters over time. Furthermore, it can also be observed that the RMSE of the MML estimator tends to converge to the LB at low noise variances for both time intervals. However, it can be noticed that the RMSE of the MML estimator is lower than the established bounds at high noise levels. This can be attributed to the fact that the location of the VLC receiver is restricted to a room with dimensions $4\,$m $\times$ $4\,$m $\times$ $3\,$m, and the MML estimator in \eqref{eq:MMLestSce1} performs its search within this specific volume; that is, $\mtL=[-2,2]\times[-2,2]\times[0,3]$ in \eqref{eq:MMLestSce1}.

\begin{figure}
 \includegraphics[width=0.95\linewidth]{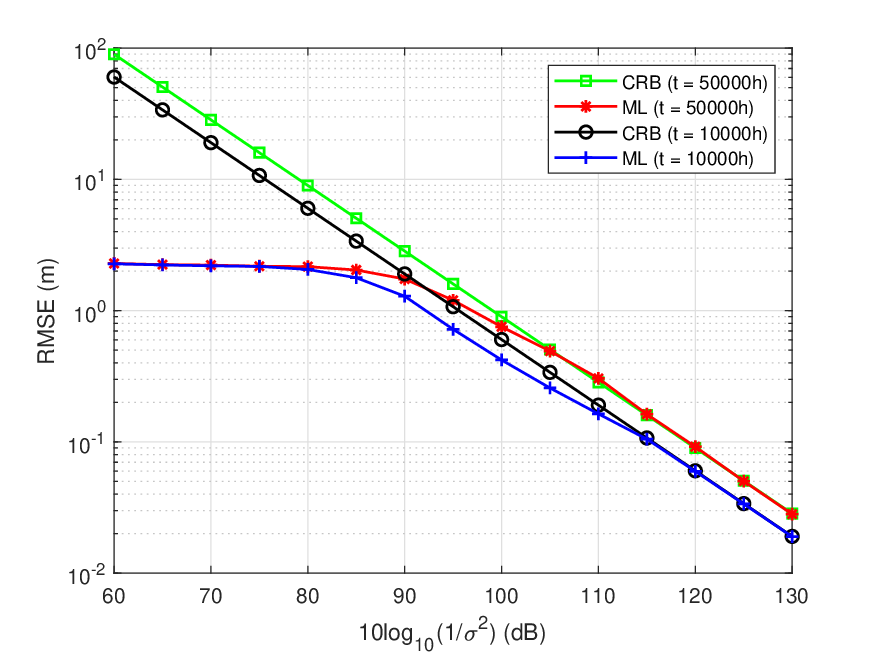}
\centering
\caption{Performance of the ML estimator and the corresponding CRB versus $10\log_{10} (1/\sigma^2)$ in Scenario~2 for LED decay rates of $\alpha = 10^{-5}$ and time intervals of $t \in \{10000\,h,\, 50000\,h\}$ when the VLC receiver is positioned at $(0.5,0,5,0.85)\,$m.}
 \label{fig:scen2figure}
\end{figure}

In Fig.~\ref{fig:scen2figure}, the RMSE performance of the ML estimator in Scenario~2 and the corresponding CRB are depicted versus $10\log_{10} (1/\sigma^2)$ for $t=10000\,$h and $t=50000\,$h. Similar to Fig.~\ref{fig:scen1figure}, it is observed that the RMSEs of the ML estimators for both time intervals are lower than the corresponding CRBs at high levels of noise variances, which is primarily due to the spatial constraints within which the ML estimator in \eqref{eq:sce2PosEst} and \eqref{eq:sce2PosEstlR} performs its search. At lower levels of noise variances, the RMSEs of the ML estimators converge to the CRBs for both time intervals. Additionally, it is important to highlight that while the RMSE values of the ML estimator approach the lower bound for both time intervals at very low noise levels, the ML estimator actually exhibits better RMSE performance at $t = 10000\,$h compared to the established lower bound, especially under conditions of moderate noise levels. This can be attributed to the constraint imposed on the decay rate estimates within the ML estimator. Namely, while the decay rate parameter is considered as a real number in the CRB calculations, it is searched among positive real numbers in the ML estimator in \eqref{eq:estSce2} (equivalently, in \eqref{eq:sce2PosEst} and \eqref{eq:sce2PosEstlR}). Furthermore, as in Scenario~1, the performance of the ML estimator for $t = 10000\,$h is better than that for $t = 50000\,$h, which is due to the power degradation over time, as discussed previously.

\begin{figure}
\includegraphics[width=0.96\linewidth]{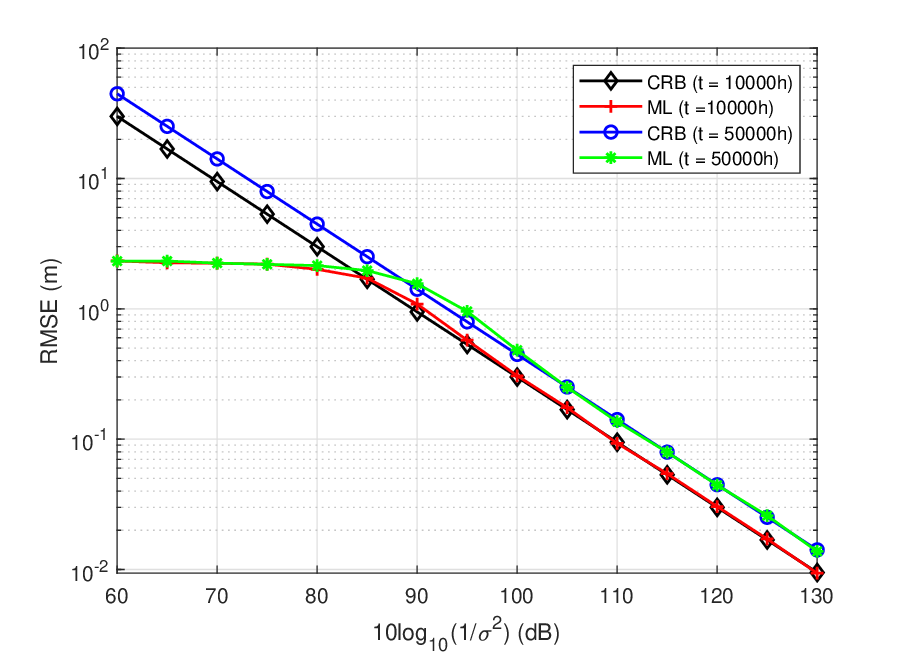}
\centering
\caption{Performance of the ML estimator and the corresponding CRB versus $10\log_{10} (1/\sigma^2)$ in Scenario~3 for LED decay rates of $\alpha = 10^{-5}$ and time intervals of $t \in \{10000\,h,\, 50000\,h\}$ when the VLC receiver is positioned at $(0.5,0,5,0.85)\,$m.}
\label{fig:scen3figure}
\end{figure}

In Fig. \ref{fig:scen3figure}, we illustrate the RMSE performance of the ML estimator and the established theoretical lower bound in Scenario~3. The estimator performance is plotted as a function of $10\log_{10} (1/\sigma^2)$ for two different time intervals. Similar to Figs.~\ref{fig:scen1figure} and \ref{fig:scen2figure}, the RMSEs of the ML estimators for both time intervals are lower than the corresponding CRBs at high levels of noise variances due to the consideration of the room dimensions in the ML estimator in \eqref{eq:MLestSce3}. Also, at lower levels of noise variances, the RMSEs of the ML estimators converge to the lower bounds for both time intervals. Again, as in Scenario~1 and Scenario~2, the performance of the ML estimator at $t = 10000\,$h is better than that at $t = 50000\,$h, which can be attributed to the gradual deterioration in power over time, as previously explained.

\begin{figure}
 \includegraphics[width=0.96\linewidth]{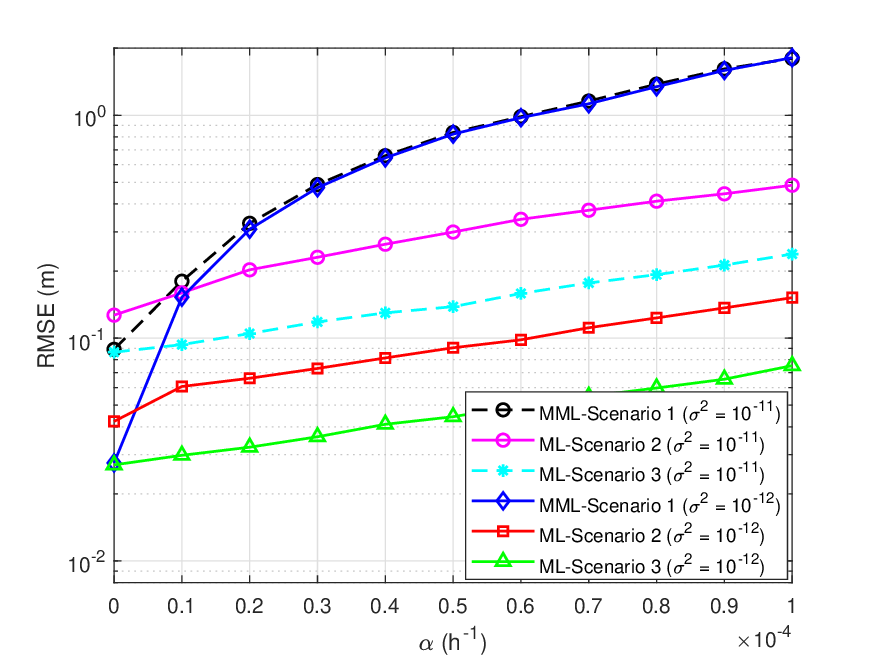}
\centering
\caption{RMSE performance of all the estimators versus decay rate, $\alpha$, for $t = 10000\,$h and the noise levels of $\sigma^2 = 10^{-11}$ and $\sigma^2 = 10^{-12}$ when the VLC receiver is positioned at $(0.5,0,5,0.85)\,$m.}
\label{fig:theoreticalbdsvsalpha}
\end{figure}

In order to compare the estimators in Scenarios 1, 2, and 3, we present in Fig.~\ref{fig:theoreticalbdsvsalpha} the RMSE performances of the estimators as a function of the decay rate $\alpha$ of the LEDs considering noise variance of $\sigma^2=10^{-11}$ and $\sigma^2=10^{-12}$. It is observed that the MML estimator under Scenario~1 and the ML estimator under Scenario~3 attain the same RMSE values for $\alpha = 0$ at both noise levels. This is due to the fact that when $\alpha = 0$, there is no mismatch between the assumed model and the true model; hence, Scenario 1 and Scenario~3 coincide. However, the ML estimator under Scenario~2 performs worse since it is unaware of the $\alpha$ value and tries estimating it. Also, it can be observed that as the decay rate $\alpha$ increases, that is, as the mismatch between the true and the assumed models increases, the RMSE performance of the  MML estimator deteriorates significantly in comparison to the RMSE performances of the ML estimators in Scenario~2 and Scenario~3. Again, it can be observed that the ML estimator under Scenario~3 always outperforms that in Scenario~2 since the $\alpha$ value is known under Scenario~3 and unknown under Scenario~2. Additionally, it is worth noting that the RMSE values of the MML estimator does not improve significantly with decreasing noise variance as $\alpha$ increases (as the model mismatch becomes the main source of error). Furthermore, it can be noted that the sensitivity to the model mismatch increases with decreasing noise variance as it is more pronounced at $\sigma^2=10^{-12}$ than at $\sigma^2=10^{-11}$. This demonstrates that being unaware of the true power decay pattern of the LEDs can be a significant inhibiting factor for accurate visible light positioning at low noise levels.  

\begin{figure}
\includegraphics[width=0.96\linewidth]{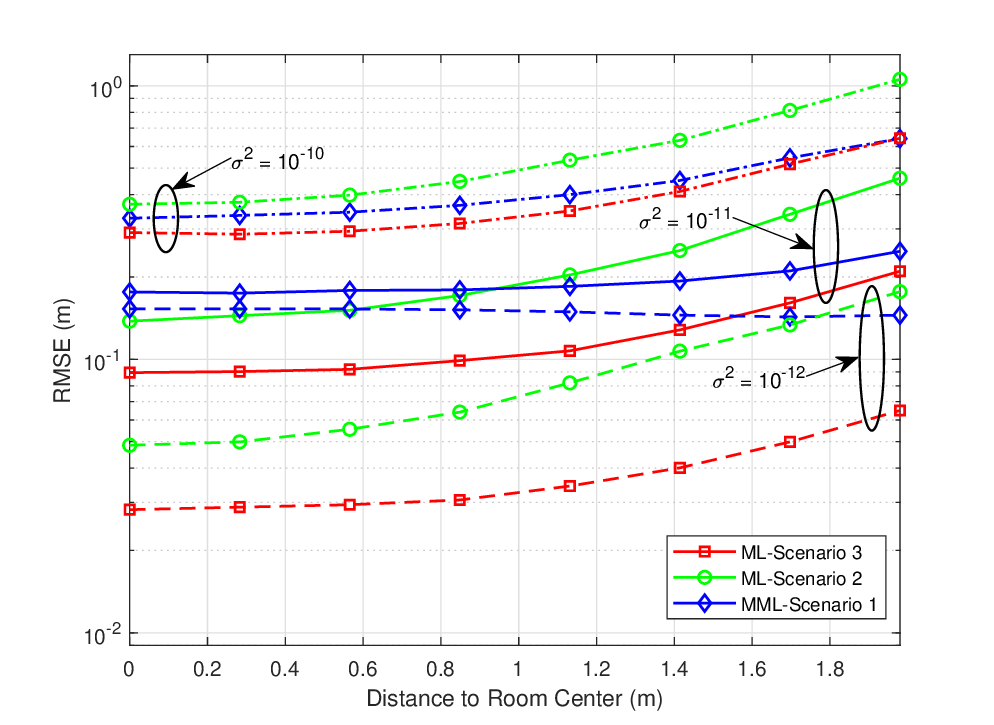}
\centering
\caption{RMSE performance of the estimators versus distance of the VLC receiver to the room center at a fixed height for LED decay rate of $\alpha = 10^{-5}$, time interval of $t = 10000\,$h, and noise variances of $\sigma^2\in\{10^{-12},10^{-11},10^{-10}\}$ when the VLC receiver moves on the straight line from $(0, 0, 0.85)\,$m towards $(1.4, 1.4, 0.85)\,$m.}
 \label{fig:estimatorsvsdistance}
\end{figure}

In Fig.~\ref{fig:estimatorsvsdistance}, we plot the RMSE performances of the estimators under the three scenarios versus the distance of the VLC receiver to the room center at a fixed height when the VLC receiver moves on the straight path from $(0, 0, 0.85)\,$m towards $(1.4, 1.4, 0.85)\,$m by considering noise variances of $\sigma^2=10^{-12}$, $\sigma^2=10^{-11}$ and $\sigma^2=10^{-10}$. It is noted from the figure that as the VLC receiver moves away from the center, the RMSEs of the estimators tend to increase for all noise levels due to decreasing power levels at the VLC receiver. An exception to this is observed for $\sigma^2=10^{-12}$ in Scenario~1 as the model mismatch becomes the main source of error for low noise variances, and the RMSE value stays almost the same. In addition, the RMSE values of the ML estimator under Scenario~3 consistently remain lower than those of the estimators in Scenario~1 and Scenario~2 for each value of the noise variance. Moreover, it is worth noting that the ML estimator under Scenario~2 achieves lower RMSE values than the MML estimator under Scenario~1 over a wider range of distances as the variance decreases.
%for the high noise variance setting (i.e., $\sigma^2=10^{-10}$), the RMSE values of the MML estimator are always lower than those of the ML estimator under Scenario~2. However, for the lower noise variances, the RMSE values of the ML estimator under Scenario~2 are lower than the RMSE values of the MML estimator while the VLC receiver moves from the center to the distance of $0.94\,$m and $1.77\,$m for the cases of $\sigma^2=10^{-11}$ and $\sigma^2=10^{-12}$, respectively. From that distance onward, the RMSE values of the ML estimator under Scenario~2 increase significantly in comparison with the RMSE values of the MML estimator. 
%These observations are mainly due to the fact that the position estimation is performed based on a mismatched model in Scenario~1 with no consideration of power decays whereas the aim is to jointly estimate the decay rate parameter $\alpha$ and the position of the VLC receiver in Scenario~2. 
This is mainly due to the fact that the effects of the model mismatch in Scenario~1 become more significant at low noise variances (as observed in Fig.~\ref{fig:scen1figure}). 
%Therefore, for high SNRs (low noise levels), the RMSE values of the MML estimator become less responsive to SNR as the model mismatch becomes the main source of error (see Fig.~\ref{fig:scen1figure}) while the RMSE values of the ML estimator under Scenario~2 are sensitive to SNR (see Fig.~\ref{fig:scen2figure}). 
%This situation is observed in Fig.~\ref{fig:estimatorsvsdistance} 
It is also noted that as the VLC receiver moves from $(0, 0, 0.85)\,$m to $(1.4, 1.4, 0.85)\,$m, the power measurements generally reduce and the ML estimator in Scenario~2 cannot estimate the decay rate and the position accurately; hence, is outperformed by the MML estimator in Scenario~1 after a distance of $0.94\,$m for  $\sigma^2=10^{-11}$ and $1.77\,$m for  $\sigma^2=10^{-12}$. In particular, since both the position and the decay rate are jointly estimated by the ML estimator in Scenario~2, its performance has the highest sensitivity to the changes in the received power levels (equivalently, to the SNRs), as noted from Fig.~\ref{fig:estimatorsvsdistance}. 
%However, as the VLC receiver moves further away, lower power measurements are obtained by the VLC receiver and the MML estimator in Scenario~1 tends to perform better than the ML estimator under Scenario~2, which tries estimating the position and the decay rate jointly. As this joint estimation becomes more challenging for the case of high noise variances, the ML estimator in Scenario~2 is outperformed by the MML estimator when $\sigma^2=10^{-10}$. 
%(Of course, if $\alpha$ gets large, the ML estimator in Scenario~2 can outperform the MML estimator, as investigated in Fig.~\ref{fig:theoreticalbdsvsalpha}.)

\section{Concluding Remarks}\label{sec:Conc}

We have investigated the received power based position estimation problem in visible light systems in the presence of luminous flux degradation of LEDs. We have  considered three different scenarios with varying levels of knowledge about the LED degradation. When the VLC receiver is unaware of the degradation and performs position estimation under that assumption, there exists a mismatch between the true model and the assumed model. In this scenario (Scenario~1), we have derived the MCRB and the MML estimator to quantify the performance loss due to this model mismatch. Then, we have assumed that the VLC receiver knows the power decay formula for the LEDs but does not know the decay rate. In this scenario (Scenario~2), we derived the CRB and the ML estimator. In the final scenario (Scenario~3), we have considered the presence of full knowledge (both the model and the decay rate parameters) at the VLC receiver, and obtained the CRB and the ML estimator. By comparing the theoretical limits and the estimators in these three scenarios, we have revealed the effects of the knowledge of the LED degradation model and the decay rate parameter on position estimation performance. In particular, the model mismatch results in significant performance loss at high SNRs, hence, it prevents achieving very low estimation errors. With the knowledge of the degradation model, this performance loss at high SNRs can significantly be reduced by performing joint estimation of the position and the decay rate parameter. Of course, when the decay rate parameter is also known beforehand, the best position estimation accuracy is achieved by performing position estimation based on the exactly known model.

\appendix
\section{Appendices}

\subsection{Derivation of \eqref{pseudoTrueParameter}}\label{app:Pseudo}

From the definition of the KL divergence, the objective function in \eqref{eq:pseudoTrue} can be stated as
\begin{align}\nonumber
D(p(\PR) \| \tilde{p}(\PR\,|\,\lR)) 
&= \int \log \left( { \frac{p(\PR)}{\tilde{p}(\PR\,|\,\lR)} } \right)p(\PR)d\PR
\\\label{eq:ExpLogRatio}
&={\mathbb{E}}_p \bigg\{\log \left( { \frac{p(\PR)}{\tilde{p}(\PR\,|\,\lR)} }\right) \bigg\}\cdot
\end{align}
Based on \eqref{eq:likeSce1} and \eqref{eq:likeSce1_2}, the expression in \eqref{eq:ExpLogRatio} can be calculated as follows:
\begin{align}\nonumber
D(p(\PR) \| \tilde{p}(\PR\,|\,\lR)) &= 
{\mathbb{E}}_p \Bigg\{
\sum_{i=1}^{\NL}\frac{(\PRi-\PTi^0 \, R_p h_i(\lR))^2}{2\sigma_i^2} -\sum_{j=1}^{\NL}\frac{(\PRj-\PTj^0 \, e^{-\alpha_j t} R_p h_j(\lR))^2}{2\sigma_j^2}
\Bigg\}
\nonumber
\\\label{eq:KLdiverFinal}
&=\sum_{i=1}^{\NL}\frac{\big(\PTi^0 \, e^{-\alpha_i t} R_p h_i(\bar{\boldsymbol{l}}_R) -\PTi^0 \, R_p h_i(\lR) \big)^2}{2\sigma_i^2} \cdot
\end{align}
Therefore, the pseudo-true parameter in \eqref{eq:pseudoTrue} can be calculated as in \eqref{pseudoTrueParameter} based on the final expression in \eqref{eq:KLdiverFinal}.

%\begin{equation}
%\lR^0 =
%\underset{\lR}{\operatorname{argmin}}\sum_{i=1}^{\NL}%\frac{(\PTi^0 \, e^{-\alpha_i t} R_p %h_i({\boldsymbol{\bar{l}}}_R) -\PTi^0 \, R_p h_i(\lR) %)^2}{2\sigma_i^2} 
%\end{equation}

\subsection{First and Second Derivatives of Channel Coefficient}\label{app:ChanCoefDer}

The first-order partial derivatives of the channel coefficient $h_i(\lR)$ with respect to the position parameter $\lR$ can be calculated via \eqref{eq:alpha} as follows:
\begin{align}
\label{eq:firstPartial}
\frac{\partial h_i(\lR) }{\partial \lR(n)} =
\frac{(\mi+1) \Ar}{2\pi} \left(
\frac{1}{v(\lR)}\frac{\partial u(\lR) }{\partial \lR(n)} 
- \frac{u(\lR)}{v^2(\lR)}\frac{\partial v(\lR) }{\partial \lR(n)}
\right)
\end{align}
where
\begin{align}\label{eq:ulr}
u(\lR) &= \left[(\lR - \lT^i)^{T} \nT^i \right]^{\mi} (\lT^i-\lR)^{T}\nR,
\\\label{eq:vlr}
v(\lR) &= \norm{\lR - \lT^i}^{\mi+3},
\\\label{eq:ulr_der}
\nonumber
\frac{\partial u(\lR) }{\partial \lR(n)} &=\mi((\lT^i-\lR)^{T}\nR)
\left[(\lR - \lT^i)^{T} \nT^i \right]^{\mi - 1}\\&~~~~\times\nT^i(n)\,
-\left[(\lR - \lT^i)^{T} \nT^i \right]^{\mi}\nR(n),
\\\label{eq:vlr_der}
\frac{\partial v(\lR) }{\partial \lR(n)}&= (\mi+3)(\lR(n)- \lT^i(n))\norm{\lR - \lT^i}^{\mi+1}.
\end{align}

In addition, the second-order partial derivatives of the channel coefficient with respect to the position parameter can be calculated from \eqref{eq:firstPartial} as follows:
\begin{align}
\label{eq:SecDeri_hi}
\nonumber
\frac{\partial^{2} h_i(\lR)}{\partial \lR(m)\partial \lR(n)} = 
\frac{(\mi+1) \Ar}{2\pi}&\Bigg(\frac{1}{v(\lR)}\frac{\partial^2 u(\lR) }{\partial \lR(m)\partial \lR(n)} -\frac{1}{v^2(\lR)}\frac{\partial v(\lR) }{\partial \lR(m)}\frac{\partial u(\lR) }{\partial \lR(n)} 
\nonumber
\\&-\frac{1}{q(\lR)}\frac{\partial b_n(\lR) }{\partial \lR(m)} +\frac{b_n(\lR)}{q^2(\lR)}\frac{\partial q(\lR) }{\partial \lR(m)}
\Bigg)
\end{align}
where
\begin{align}
\label{}
\nonumber
&\frac{\partial^2 u(\lR) }{\partial \lR(m)\partial \lR(n)} = \,\mi(\mi-1)((\lT^i-\lR)^{T}\nR)\left[(\lR - \lT^i)^{T} \nT^i \right]^{\mi - 2}\nT^i(n)\nT^i(m)\\&  
{\hspace{3cm}}-\mi\left[(\lR - \lT^i)^{T} \nT^i \right]^{\mi - 1}(\nT^i(n)\nR(m) +\nT^i(m)\nR(n)),
\\\label{}
&b_n(\lR) \triangleq  \,(\mi+3)(\lR(n)- \lT^i(n))\left[(\lR - \lT^i)^{T} \nT^i \right]^{\mi}((\lT^i-\lR)^{T}\nR),
\\
&q(\lR) \triangleq  \norm{\lR- \lT^i }^{\mi+5},
\\\label{}
&\frac{\partial b_n(\lR) }{\partial \lR(m)} =
\begin{cases}\small
(\mi+3)(\lR(n)- \lT^i(n))[\mi\nT(m)
\small((\lT^i-\lR)^{T}\nR)\small\left((\lR - \lT^i)^{T} \nT^i \right)^{\mi-1} \\\small-\nR(m)\left((\lR - \lT^i)^{T} \nT^i \right)^{\mi} ],
~{\rm{if}}~m\neq n
\\\small
(\mi + 3)\left[(\lR - \lT^i)^{T} \nT^i \right]^{\mi}((\lT^i-\lR)^{T}\nR)+ (\mi+3)(\lR(m)- \lT^i(m)) \\\small\times[\mi\nT(m)((\lT^i-\lR)^{T}\nR)
\left((\lR - \lT^i)^{T} \nT^i \right)^{\mi - 1}\small-\nR(m)\left((\lR - \lT^i)^{T} \nT^i \right)^{\mi} ],
~{\rm{if}}~m= n
\end{cases}\normalsize
\\\label{eq:der_q}
&\frac{\partial q(\lR) }{\partial \lR(m)} = (\mi+5)(\lR(m)- \lT^i(m))\norm{\lR - \lT^i}^{\mi+3}.
\end{align}
Although the first-order derivative of $h_i(\lR)$ was presented in \cite[Eqn.~(14)]{Direct_TCOM}, the second-order derivatives given by \eqref{eq:SecDeri_hi}--\eqref{eq:der_q} were not available in the literature.

\bibliographystyle{IEEEtran}

\bibliography{Ref_proc_IEEE}

%\begin{IEEEbiography}{Issifu Iddrisu}{\space}
%\end{IEEEbiography}

%\begin{IEEEbiography}{Sinan Gezici}
%{\space}
%\end{IEEEbiography}

\end{document}